\documentclass{aa}
\usepackage{graphics,latexsym,amssymb,times,psfig,amsmath}
\def\gsim{ \lower .75ex \hbox{$\sim$} \llap{\raise .27ex \hbox{$>$}} } 
\def\lsim{ \lower .75ex\hbox{$\sim$} \llap{\raise .27ex \hbox{$<$}} }

\begin{document}
   
   \title{On the interpretation of the spectral--energy correlations
          in long Gamma--Ray Bursts
         }

\author{Nava L. \inst{1,2}, 
        Ghisellini G. \inst{1},
        Ghirlanda G.  \inst{1},
	Tavecchio F.  \inst{1},
        Firmani C.    \inst{1,3}
	  }

   \offprints{Gabriele Ghisellini \\gabriele@merate.mi.astro.it}

\institute{
Osservatorio Astronomico di Brera, via Bianchi 46, Merate  Italy  \and 
Univ. di Milano--Bicocca, P.za della Scienza 3, I--20126, Milano, Italy. \and
Instituto de Astronom\'{\i}a, U.N.A.M., A.P. 70-264, 04510, M\'exico, D.F., M\'exico
             }

\date{Received ... / Accepted ...}

\titlerunning{Spectral--energy correlations in long GRBs}
\authorrunning{L. Nava et al.}

\abstract{ Recently, Liang \& Zhang (2005) found a tight correlation
involving only observable quantities, namely the isotropic emitted
energy $E_{\rm \gamma, iso}$, the energy of the peak of the prompt
spectrum $E^\prime_{\rm p}$, and the jet break time $t^\prime_{\rm
j}$.  This phenomenological correlation can have a first explanation
in the framework of jetted fireballs, whose semiaperture angle
$\theta_{\rm j}$ is indeed measured by the jet break time
$t^\prime_{\rm j}$.  By correcting $E_{\rm \gamma, iso}$ for the angle
$\theta_{\rm j}$ one obtains the so called Ghirlanda correlation which
links the collimation corrected energy $E_\gamma$ and $E^\prime_{\rm
p}$.  There are two ways to derive $\theta_{\rm j}$ from
$t^\prime_{\rm j}$ in the ``standard'' scenario, corresponding to an
homogeneous or instead to a wind--like circumburst medium.  We show
that the Ghirlanda correlation with a wind--like circumburst medium is
as tight as (if not tighter) than the Ghirlanda correlation found in
the case of an homogeneous medium.  There are therefore two Ghirlanda
correlations, both entirely consistent with the phenomenological Liang
\& Zhang relation.  We then suggest to consider the difference between
the {\it observed} correlations and the ones one would see in the {\it
comoving frame} (i.e. moving with the same bulk Lorentz factor of the
fireball).  Since both $E_{\rm p}$ and $E_\gamma$ transform in the
same way, the wind--like Ghirlanda relation, which is linear, remains
linear also in the comoving frame, no matter the distribution of bulk
Lorentz factors.  Instead, in the homogeneous density case, one is
forced to assume the existence of a strict relation between the bulk
Lorentz factor and the total energy, which in turn put constraints on
the radiation mechanisms of the prompt emission.  The wind--like
Ghirlanda correlation, being linear, corresponds to different bursts
having the same number of photons.  \keywords{Gamma rays: bursts ---
Radiation mechanisms: non-thermal --- X--rays: general } }
             
\maketitle

\section{Introduction}

The correlation between the collimation corrected emitted energy in
the prompt phase of GRBs ($E_\gamma$) and the peak energy of the
$\nu E_\nu$ prompt spectrum ($E_{\rm p}$), the so--called
``Ghirlanda" relation, is the fundamental tool for the
cosmological use of GRBs.

Its importance calls for a robust and convincing explanation.
Possible ideas and interpretations have already been proposed, but
requiring some ad hoc assumptions: different authors (Eichler \&
Levinson 2004; Levinson \& Eichler 2005; Yamazaki, Ioka \& Nakamura
2004; Toma, Yamazaki \& Nakamura 2005) have underlined the importance
of viewing angle effects assuming different geometries of the fireball
(annular or patchy); while Rees \& Meszaros (2005) pointed out that a
strict relation between total energy and typical peak frequency can be
understood in a easier way if the underlying emission process is
thermal, therefore suggesting that dissipative processes in the
photosphere of the fireball can increase the thermal (black--body
like) photon content of the fireball itself.

The collimation corrected energy $E_\gamma$ is derived by multiplying
the isotropic emitted energy $E_{\rm \gamma, iso}$ by the factor
$(1-\cos\theta_{\rm j})$, where $\theta_{\rm j}$ is the semiaperture
angle of the jet.  To derive it, one must assume the uniform jet model
and the density profile of the circumburst medium (e.g. homogeneous or
wind--like).

The Ghirlanda correlation has been derived for an homogeneous density
(i.e.  $n$ constant) and there are a few critical points concerning
its derivation which are now under discussion.

Firstly, in this scenario the jet opening angle is (Sari et al. 1999)
\begin{equation} 
\theta_{\rm j}=0.161 \left({ t_{\rm j,d} \over 1+z}\right)^{3/8} 
\left({n \, \eta_{\gamma}\over E_{\rm \gamma,iso,52}}\right)^{1/8} 
\label{theta} 
\end{equation} 
where $z$ is the redshift, $\eta_\gamma$ is the radiative efficiency
and $t_{\rm j,d}$ is the achromatic break time, measured in days, of
the afterglow lightcurve \footnote{Here we adopt the notation
$Q=10^xQ_x$, and use cgs units unless otherwise noted.}.  The
efficiency $\eta_\gamma$ relates the isotropic kinetic energy of the
fireball after the prompt phase, $E_{\rm k, iso}$, to the prompt
emitted energy $E_{\rm \gamma, iso}$, through $E_{\rm k, iso}= E_{\rm
\gamma, iso}/\eta_\gamma$. This implicitly assumes that $\eta_\gamma
\ll 1$ otherwise the remaining kinetic energy after the prompt
emission is instead $E_{\rm k, iso}=E_{\rm \gamma, iso}
(1-\eta_\gamma)/\eta_\gamma$.  This efficiency, in principle, could be
different from bursts to bursts, but in the absence of any hints of
how its value changes as a function of other properties of the bursts
and favoured by its low power in Eq.1, one assumes a constant value
for all bursts, i.e. $\eta_\gamma =0.2$ (after its first use by Frail
et al. 2001, following the estimate of this parameter in GRB 970508).

Secondly, the density of the circumburst medium $n$ can in principle
be estimated through accurate fits to the lightcurves at different
frequencies [or, equivalently, fitting the spectral energy
distributions (SEDs) at different times], but the fact that the
emitted synchrotron spectrum is insensitive to $n$ in the regime of
fast cooling makes the $n$ estimates somewhat uncertain.  Furthermore,
only for a minority of bursts we have enough data to constrain $n$
even in a rather poor way (see e.g. Panaitescu \& Kumar 2000, 2001).
From this partial information, however, the estimated values of $n$
range from $\sim 0.1$ to $\sim 30$ cm$^{-3}$ with a clear preference
for an homogeneous density scenario.  Wind density profiles are
acceptable and even preferred in few cases, in which however one
cannot exclude the homogeneous density case.  This led to the choice,
made by Ghirlanda, Ghisellini \& Lazzati (2004, hereafter GGL04) to
assume an homogeneous density scenario for all bursts and assign, to
all cases in which $n$ was not estimated, a range of possible values
of $n$, from 1 to 10 cm$^{-3}$, to derive the value of $\theta_{\rm
j}$ using Eq. \ref{theta}, and, more importantly, the associated
error.  However, if GRBs are expected to originate from the death of
very massive stars, the wind density profile appears as the most
natural outcome of the final stages of the evolution of the burst
progenitor.

These issues on one hand must caution us about the possible scatter of
the points in the Ghirlanda correlation, which may well be the results
of bursts having a distribution (instead of a single value) of $n$ and
$\eta_\gamma$, but on the other hand the extremely small scatter found
suggests that, for the considered bursts, these values are indeed
clustered in a small range.

In any case, these concerns (i.e. a possible distribution of $n$ and
$\eta_\gamma$ values) have been completely overcome by the finding, by
Liang \& Zhang (2005, hereafter LZ05), of a phenomenological and
model--independent correlation between $E_{\rm \gamma, iso}$,
$E^\prime_{\rm p}$ and $t^\prime_{\rm j}$.  By considering 15 GRBs,
and a flat Universe cosmology with $\Omega_{\rm M}=0.28$ and
$h_0=0.713$, the correlation found by LZ05 takes the form:
\begin{equation}
E_{\rm \gamma,iso,52}=(0.85\pm 0.21)
\left( {E^\prime_{\rm p} \over 100~{\rm keV} }\right)^{1.94\pm 0.17}
t_{\rm j,d}^{\prime -1.24\pm 0.23}
\label{lz}
\end{equation}
where primed quantities are calculated in the rest frame of the
GRB, i.e. $t^\prime=t/(1+z)$ and $E^\prime_{\rm p}= E_{\rm p}(1+z)$.  
The scatter of the data points around this correlation is
small enough to enable LZ05 to use it for finding constraints to the
cosmological parameters.

The main aim of the present paper is to discuss a few steps which we
think necessary in order to deepen our understanding of the
spectral/energy correlations in GRB, even if we do not claim to arrive
to a complete or satisfactory interpretation.  The Ghirlanda
correlation (although derived earlier than the LZ05 relation) should
be considered as a first step towards the explanation of the purely
phenomenological LZ05 relation.

The first step is to demonstrate that the LZ05 correlation is
equivalent to the Ghirlanda correlation.  Secondly, we explore what
happens to the Ghirlanda correlation if, instead of an homogeneous
medium, we assume that the density is distributed with a $r^{-2}$ wind
profile.  We demonstrate that the LZ05 is again consistent with this
new Ghirlanda--wind correlation.

We then have not one, but two possibilities of relating $E_{\rm p}$
and $E_{\gamma}$, both consistent with the LZ05 correlation.  We argue
that the new (wind--like) Ghirlanda correlation cannot be easily
discarded on the basis of the fit to the afterglow SEDs, which prefer
the uniform density case. The present small sample of GRBs does not
allow to test the two possibilities, but we mention one test to do so
when estimates of the initial bulk Lorentz factor will be available.

We then discuss a third step, pointing out that the observed $E_{\rm
p}$ and $E_{\gamma}$ are affected by the relativistic motion of the
fireball.  In the simplest and standard scenario, which assumes that
the observer's line of sight is within the jet opening angle and that
the jet is homogeneous, $E_{\rm p}$ and $E_{\gamma}$ are both boosted
by a factor $\sim 2\Gamma$,  where $\Gamma$ is the bulk Lorentz
factor of the fireball.  Therefore what we see is an {\it apparent}
correlation.  Imposing that some correlation (even different from the
observed one) exists in the comoving frame allows to put interesting
constraints on both the dynamics and the emission processes of bursts.

As part of our effort, we report the updated tables for the relevant
parameters used in deriving all the results and correlations discussed
in the present paper.  We discuss case--by--case the differences and
changes of these quantities with respect to previously published
papers.

\begin{table*}
\begin{center}
\begin{tabular}{lllllllllll}
\\
\hline
\hline
\\
GRB &$z$ &$\alpha$ &$\beta$ &Fluence    &Range &$E_{\rm p}$ &Ref$^a$ &$t_{\rm j}$ &$n$        &Ref$^b$ \\ 
        &      &         &        &erg/cm$^2$ &keV   &keV         &        &days        &cm$^{-3}$  & \\
\\
\hline
\\ 
970828  & 0.958 &--0.70 [0.08]  &--2.07 [0.37] &9.6e--5 [0.9] & 20--2000 &298 [60] &1, 2   &2.2  (0.4)  &3.0 [2.76] &25, ... \\ 
980703  & 0.966 &--1.31 [0.14]  &--2.40 [0.26] &2.26e--5[0.23]& 20--2000 &254 [51] &3, 2   &3.4  (0.5)  &28  (10)   &25, 25  \\  
990123  & 1.600 &--0.89 (0.08)  &--2.45 (0.97) &3.0e--4 (0.4) & 40--700  &781 (62) &4, 5   &2.04 (0.46) &3.0 [2.76] &26, ... \\ 
990510  & 1.619 &--1.23 (0.05)  &--2.7 (0.4)   &1.9e--5 (0.2) & 40--700  &161 (16) &6, 5   &1.6  (0.2)  &0.29 (0.13)&27, 36 \\ 
990705  & 0.843 &--1.05 (0.21)  &--2.2 (0.1)   &7.5e--5 (0.8) & 40--700  &189 (15) &7, 5   &1.0  (0.2)  &3.0 [2.76] &25, ... \\ 
990712  & 0.433 &--1.88 (0.07)  &--2.48 (0.56) &6.5e--6 (0.3) & 40--700  &65 (11)  &6, 5   &1.6  (0.2)  &3.0 [2.76] &28, ...  \\ 
991216  & 1.02  &--1.23 [0.13]  &--2.18 [0.39] &1.9e--4 [0.2] & 20--2000 &318 [64] &8, 2   &1.2  (0.4)  &4.7 (3.5)  &25, 36 \\ 
011211  & 2.140 &--0.84 (0.09)  &...           &2.6e--6 [0.3] & 40--700  &59 (8)   &9, 9   &1.56 [0.16] &3.0 [2.76] &29, ... \\ 
020124  & 3.198 &--0.87 (0.17)  &--2.6  (0.65) &8.1e--6 (0.9) & 2--400   &93 (27)  &10, 11 &3.0  [0.4]  &3.0 [2.76] &30, ... \\ 
020405  & 0.695 &--0.0 (0.25)   &--1.87 (0.23) &7.4e--5 [0.7] & 15--2000 &364 (101)&12, 12 &1.67 (0.52) &3.0 [2.76] &12, ...\\    
020813  & 1.255 &--1.05 [0.11]  &...           &1.0e--4 [0.1] & 30--400  &212 [42] &13, 14 &0.43 (0.06) &3.0 [2.76] &25, ... \\ 
021004  & 2.335 &--1.0  (0.2)   &...           &2.6e--6 (0.6) & 2--400   &80 (35)  &15, 16 &4.74 [0.5]  &3.0 [2.76] &31, ...\\
030226  & 1.986 &--0.9 (0.2)    &...           &5.6e--6 (0.6) & 2--400   &97 (21)  &17, 16 &1.04 (0.12) &3.0 [2.76] &32, ...\\ 
030328  & 1.520 &--1.14 (0.03)  &--2.1 (0.3)   &3.7e--5 (0.14)& 2--400   &130 (14) &18, 16 &0.8  [0.1]  &3.0 [2.76] &33, ...\\ 
030329  & 0.169 &--1.32 (0.02)  &--2.44 (0.08) &1.2e--4 [0.12]& 30--400  &70 (2)   &19, 20 &0.5  (0.1)  &2.2 [0.80] &34, 38 \\    
030429  & 2.656 &--1.1  (0.3)   &...           &8.5e--7 (1.4) & 2--400   &35 (10)  &21, 16 &1.77 (1.0)  &3.0 [2.76] &35, ...\\ 
041006  & 0.716 &--1.37 [0.14]  &...           &2.0e--5 [0.2] & 25--100  &63 [13]  &22, 22 &0.16 (0.04) &3.0 [2.76] &37, ...\\
050525  & 0.606 &--0.99 (0.11)  &...           &2.01e--5 (0.05)& 15--350 &79 (3.5) &23, 24 &0.28 (0.12) &3.0 [2.76] &39, ...\\
\\
\hline
\end{tabular}
\end{center}
\caption{ Input parameters for the bursts of our sample. $\alpha$ and
$\beta$ are the photon spectral indices of the prompt emission
spectrum and $E_{\rm p}$ represents the (observed) peak energy of the
$\nu F_{\nu}$ spectrum. When the errors are not given in the original
reference, we assumed an average error (values in square brackets),
otherwise we list the originally reported error (in round brackets).
$^a$References are given in order for the redshift ($z$) and for the
spectral parameters ($\alpha$, $\beta$, fluence and its energy
interval): 1 Djorgovsky et al.  2001; 2 Jimenez et al.  2001; 3
Djorgovsky et al.  1998; 4 Hjorth et al.  1999; 5 Amati et al.  2002;
6 Vreeswijk et al.  2001; 7 Amati et al.  2000; 8 Vreeswijk et al.
1999; 9 Amati 2004; 10 Hjorth et al. 2003; 11 Atteia et al. 2005; 12
Price et al.  2003; 13 Barth et al. 2003; 14 Barraud et al.  2003; 15
Moller et al.  2002; 16 Sakamoto et al.  2005; 17 Greiner et al.
2003a; 18 Rol et al.  2003; 19 Greiner et al.  2003b; 20 Vanderspek et
al.  2004; 21 Weidinger et al.  2003; 22
http://space.mit.edu/HETE/Bursts/; 23 Foley et al. 2005; 24 Blustin et
al. 2005; $^b$References are given in order for the observed jet break
time $t_{\rm j}$ and for the density $n$ when present: 25 Bloom et
al. 2003; 26 Kulkarni et al. 1999; 27 Israel et al.  1999; 28
Bjornsson et al. 2001; 29 Jakobsson et al.  2003; 30 Berger et
al. 2002; 31 Holland et al. 2003; 32 Klose et al. 2004; 33 Andersen et
al. 2003; 34 Berger et al. 2003; 35 Jakobsson et al. 2004; 36
Panaitescu \& Kumar 2002; 37 Stanek et al. 2005; 38 Frail et al. 2005.
39 Blustin et al. 2005 }
\label{tabin}
\end{table*}


\section{The sample}

LZ05 found their correlation (Eq. \ref{lz}) by using 15 bursts.  Most
of them are the same bursts used by GGL04, but there are some
differences in the reported parameters for the common GRBs.  Also,
since the publication of GGL04 some (primarily spectral) parameters
have been updated (as a consequence of refined analysis) and published
in the literature.  Finally, other bursts have been detected recently,
which bring the total number of GRBs with ``useful" data (i.e. $z$,
$E_{\rm p}$ and $t_{\rm j}$, excluding upper/lower limits) to 18 (at
the time of writing, September 2005).  We stress that our sample
includes only those bursts with secure measurements of $z, E_{\rm
peak},t_{\rm j}$.  Consistency checks of the upper/lower limits can be
performed (as shown in GGL04).  For these reasons we present in
Tab. \ref{tabin} our sample of 18 GRBs with the relevant input data
and corresponding references which represents the most updated
collection of published parameters which are relevant for our
analysis. We detail in the following all the relevant differences of
this table with respect to that reported in GGL04 and LZ05.

Tab. \ref{tabin} contains all GRBs reported by LZ05 with the exception
of GRB 021211, and with the addition of 4 GRBs (GRB 970828, GRB
990705, GRB 041006, GRB 050525).  Our exclusion of GRB 021211 is
motivated by the extremely uncertain determination of $t_{\rm j}$ due
to the scarcity of afterglow data and to the likely ``contamination"
of a supernova (Della Valle et al. 2003).  The value of $t_{\rm j}$
reported by LZ05 (1.4 day, see also Holland et al. 2004) should be
considered only as a lower limit to $t_{\rm j}$ (as assumed in GGL04).

For GRBs listed in Tab. \ref{tabin} which are also present in LZ04 we
have the following (minor) differences:
\begin{itemize}

\item GRB 011211: the value of $t_{\rm j}$ reported by Jakobsson et
al. (2003) is $t_{\rm j}=1.56\pm 0.02$ days.  This has been derived by
fitting the light curve with a broken power law, which very likely
severely underestimates the error on this quantity.  Note that the
$\chi^2$ of the original fit is large.  Differently from LZ05, who
used the above error, we have set the error to 10\% (i.e. the average
error on the observed jet break time) of the value of $t_{\rm j}$.

\item 
GRB 020124: the spectral parameters have been updated by Atteia et
al. (2005) with respect to the ones assumed by LZ05 from the Sakamoto
et al. (2005).

\item
GRB 020405: for the value of $E_{\rm p}$ listed in Tab. \ref{tabin} we
assume that in the original reference (Price et al. 2003), the authors
presented the value $E_0$ of the Band spectrum. 
This corresponds
to the e--folding energy of the exponential rollover of this model and
it allows to derive the peak energy as $E_{\rm p} = E_0 (\alpha + 2)$
(where $\alpha$ is the photon spectral index of the low energy power
law component of the Band model).

Note also that if $\beta=-1.87\pm0.23$, as reported in Price et
al. 2003, the spectral peak energy is unconstrained within the
observational energy band. However, the reported uncertainty (0.23)
makes the high energy spectral component consistent with
$\beta<-2$. For this reason we assumed the lowest value,
i.e.$\beta=-2.1$, to compute the isotropic equivalent energy of this
burst.

\item GRB 020813: we take the spectral parameters of this GRB from
 Barraud et al. (2003), while LZ05 use the values reported in the
 Sakamoto et al. (2005).  The reason of our choice is in the
 fact that, with the $\beta=-1.57\pm 0.04$ reported by Sakamoto et
 al. (2005), one could not define the value of $E_{\rm p}=140\pm 14$
 keV (which instead requires at least $\beta<-2$), which is reported
 in Sakamoto et al. (2005).

\item GRB 021004: the value of $t_{\rm j}=4.74$ days comes from
  Holland et al. (2003), and it is the same (as well as the quoted
  reference) of LZ05. What is different is the error: Holland et
  al. (2003) report $t_{\rm j}=4.74^{+0.14}_{-0.80}$ days.  LZ05 take
  0.14 days as the error in this quantity, while we take 0.5 days to
  better approximate the (asymmetric) error.

\item GRB 030329: We have updated the spectral parameters of this
bursts according to the published paper by Vanderspek et al. 2004
(which are somewhat different from those presented in the preprint
version of the same article).

\item GRB 050525: the prompt emission and the early afterglow detected
by the Swift BAT, UVOT and XRT instruments have been recently analyzed
by Blustin et al. (2005).  In this paper the authors suggest the
presence of a jet break in the very early afterglow lightcurve,
i.e. $t_{\rm j}=$0.16 or 0.2 days according to two different model fit
(the second value is found including also the data of Klotz et
al. 2005, while the first value refers to the Swift data only).  On
the other hand, Mirabal et al. (2005) noted a jet break time at 0.4
days after trigger, based on a large collection of data (120 frames,
still unpublished) taken with the 2.4 MDM meter telescope.  Note also
that the fit leading to $t_{\rm j}=0.2$ days assumes that the optical
afterglow has a ``jump" in its flux, but that thereafter continues to
decay normally.  In other words, this ``discontinuity" is not treated
as a ``bump" in the lightcurve (as often seen in other bursts) which
would have implied that the afterglow is composed by two contributions
(the normal afterglow plus the re-brightening component).  Fitting
with this model would result in a larger value of $t_{\rm j}$.  While
waiting for a joint fit with all the data available, and also with
different models, we decided to use $t_{\rm j}=0.28\pm 0.12$ days:
this value is intermediate between the Blustin et al. values and the
Mirabal et al. value, with an error which encompasses all values. 

\end{itemize}

\begin{table*}
\begin{center}
\begin{tabular}{llllllll}
\\
\hline
\hline
\\
GRB  & $E^\prime_{\rm p}$ & $E_{\rm \gamma, iso}$  & $\theta_{\rm j}$ & $E_{\gamma}$ & $E_{\gamma, \rm n=3}$ & $\theta_{\rm j,w}$ & $E_{\gamma,\rm w}$ \\ 
     &  keV               & erg                    & deg              & erg          & erg                & deg              & erg   \\
\\
\hline
\\ 
970828 & 583 [117]  & 2.96e53 [0.35] & 5.91  $\pm$ 0.79 & 1.57e51 (0.46) & ...            & 3.40 (0.19) & 5.21e50 (0.84) \\ 
980703 & 499 [100]  & 6.9e52  (0.82) & 11.02 $\pm$ 0.80 & 1.27e51 (0.24) & 7.29e50 (1.20) & 5.45 (0.26) & 3.12e50 (0.48) \\ 
990123 & 2031 (161) & 2.39e54 (0.28) & 3.98  $\pm$ 0.57 & 5.76e51 (1.78) & ...            & 1.84 (0.12) & 1.24e51 (0.22) \\ 
990510 & 422  (42)  & 1.78e53 [0.19] & 3.74  $\pm$ 0.28 & 3.80e50 (0.69) & 6.80e50 (0.98) & 3.31 (0.14) & 2.98e50 (0.40) \\ 
990705 & 348  (28)  & 1.82e53 (0.23) & 4.78  $\pm$ 0.66 & 6.33e50 (1.92) & ...            & 3.20 (0.19) & 2.84e50 (0.50) \\ 
990712 &  93  (16)  & 6.72e51 (1.29) & 9.47  $\pm$ 1.20 & 9.16e49 (2.90) & ...            & 8.75 (0.51) & 7.82e49 (1.76) \\ 
991216 & 642 [129]  & 6.75e53 [0.81] & 4.44  $\pm$ 0.70 & 2.03e51 (6.79) & 1.81e51 (0.50) & 2.36 (0.21) & 5.72e50 (1.24) \\ 
011211 & 185  (25)  & 5.4e52  (0.6)  & 5.38  $\pm$ 0.66 & 2.38e50 (0.64) & ...            & 4.24 (0.16) & 1.48e50 (0.20) \\ 
020124 & 390 [113]  & 2.61e53 (0.18) & 5.07  $\pm$ 0.64 & 1.02e51 (0.27) & ...            & 3.13 (0.12) & 3.90e50 (0.40) \\ 
020405 & 617 [124]  & 1.25e53 [0.13] & 6.27  $\pm$ 1.03 & 7.48e50 (2.58) & ...            & 4.08 (0.34) & 3.17e50 (0.63) \\ 
020813 & 478  [95]  & 5.78e53 [0.58] & 2.8   $\pm$ 0.36 & 6.89e50 (1.88) & ...            & 1.85 (0.08) & 3.00e50 (0.40) \\ 
021004 & 267 (117)  & 3.38e52 (0.78) & 8.47  $\pm$ 1.06 & 3.69e50 (1.25) & ...            & 6.20 (0.40) & 1.98e50 (0.52) \\
030226 & 290  [63]  & 5.43e52 (0.68) & 4.71  $\pm$ 0.58 & 1.84e50 (0.51) & ...            & 3.88 (0.17) & 1.24e50 (0.19) \\ 
030328 & 318  [33]  & 3.68e53 [0.37] & 3.58  $\pm$ 0.45 & 7.18e50 (1.93) & ...            & 2.35 (0.10) & 3.09e50 (0.40) \\ 
030329 &  82   (2)  & 1.62e52 [0.16] & 5.69  $\pm$ 0.50 & 7.99e49 (1.62) & 8.63e49 (1.56) & 5.52 (0.31) & 7.51e49 (1.13) \\ 
030429 & 128  [26]  & 1.42e52 (0.23) & 6.3   $\pm$ 1.52 & 8.57e49 (4.37) & ...            & 5.88 (0.88) & 7.48e49 (2.55) \\ 
041006 & 108  [22]  & 6.92e52 [0.7]  & 2.79  $\pm$ 0.41 & 8.18e49 (2.57) & ...            & 2.62 (0.18) & 7.25e49 (1.23) \\
050525 & 135   [8]  & 2.3e52  [0.3]  & 4.04  $\pm$ 0.8  & 5.73e49 (2.34) & ...            & 4.04 (0.45) & 5.72e49 (1.40) \\
\\
\hline
\\
\end{tabular}
\end{center}
\caption{ The rest frame peak energy $E^\prime_{\rm p}$, the
collimation corrected energy $E_{\rm \gamma, iso}$ are calculated with
the parameters reported in Tab. \ref{tabin}. The semiaperture angle
$\theta_{\rm j}$ is calculated with Eq. \ref{theta} in the case of an
homogeneous medium and the collimation corrected energy $E_{\rm
\gamma}$ is reported. We also report the values of $E_{\rm
\gamma,n=3}$ assuming $n=3$ cm$^{-3}$ for the 4 GRBs which have a
different estimate of $n$ as reported in Tab. \ref{tabin}. In the case
of a wind medium the semiaperture angle $\theta_{\rm j, w}$
(calculated by Eq. \ref{thetaw}) and the corresponding $E_{\rm \gamma,
w}$ are reported.  When the errors in $E_{\rm p}$ ($E_{\rm \gamma,
iso}$) are not given in the original reference, we assume the average
error of 20\% (11\%) (values in square brackets), otherwise we list
the originally reported error (in round brackets).  }
\label{tabout}
\end{table*}

In Tab. \ref{tabout} we report the values of the rest frame peak
energy $E_{\rm p}^\prime$ and of the isotropic equivalent energy
$E_{\gamma, \rm iso}$. 
The values of $E_{\rm \gamma, iso}$ have been
taken directly from Amati et al. (2002) in the case of GRBs detected
by $Beppo$Sax and listed in that paper, but we converted these
values to our cosmology (i.e. we use $h=0.7$ while Amati et al. 2002
used $h=0.65$).

Tab. \ref{tabin} and Tab. \ref{tabout} are the updated version of the
tables presented in GGL04 which were composed by 15 ``usable'' GRBs.
With respect to that paper, our present sample of 18 GRBs comprises 3
new GRBs (GRB 021004, GRB 041006 and GRB 050525) for which $z$,
$E_{\rm p}$ and $t_{\rm j}$ have been published and some of the
parameters have changed due to updates which appeared in the
literature.

\begin{itemize}

\item GRB 991216: the value of the density $n=4.70^{+6.8}_{-1.8}$ was
estimated by Panaitescu \& Kumar (2002). We have here assumed a
symmetric error equal to the logarithmic average, and taken $n=4.7\pm
3.5$.  This differs slightly from what assumed in GGL04 
($n=4.7\pm 2.3$).

\item GRB 011211: we changed the reference for the spectral
parameters, which is Amati (2004), and not Amati et al. 2002 as given
in GGL04.  We recalculated the value of $E_{\rm \gamma, iso}$ with
$h=0.7$.

\item GRB 020124: we updated the spectral parameters, now taken from
Atteia et al. (2005).  The main difference concerns $E_{\rm p}=120$
keV, instead of the value of 93 keV reported in GGL04.

\item GRB 030226 and GRB 030328: we now use the spectral parameters
reported in Sakamoto et al. (2005) (instead of Atteia 2003) and the
recalculated $E_{\gamma, \rm iso}$.

\item GRB 030329: as mentioned above, we updated the spectral
parameters according to the published paper by Vanderspek et al. 2004.
The listed value of $n=2.2$ cm$^{-3}$ comes from Frail et al. 2005
(GGL04, instead, quoted $n=1$ cm$^{-3}$): the associated error
encompasses the three possible values listed in Tab. 2 of Frail et
al. 2005.

\item GRB 030429: we updated the spectral parameters with Sakamoto et
al. (2005) instead of those assumed in GGL04 which were taken from the
Hete-2 web page.

\end{itemize}

\section{The spectral--energy correlations}

With the updated sample of 18 GRBs reported in Tab. \ref{tabin} we
first refit the empirical LZ05 correlation among $E_{\rm \gamma,
iso}$, $E^\prime_{\rm p}$ and $t^\prime_{\rm j}$.  We also give the
updated version of the Ghirlanda correlation (GGL04 and Ghirlanda et
al. 2005) in the case of a homogeneous density profile.  Finally we
present the Ghirlanda correlation in the case of a wind density
profile.

\subsection{The Liang--Zhang correlation revisited}

The method used by LZ05 to find the correlation between $E_{\rm
\gamma, iso}$, $E^\prime_{\rm p}$ and $t^\prime_{\rm j}$ is a
multivariate linear regression.  The significance of the multivariate
regression is estimated through the F--test and through the Spearman
$r_{\rm s}$ coefficient between $\log E_{\rm \gamma,iso}$ calculated
through Eq. \ref{lz} and the same quantity directly calculated through
\begin{equation}
E_{\rm \gamma,iso}\, =\, {4\pi d^2_L S_{\gamma} k \over 1+z}
\label{eiso}
\end{equation}
where $d_L$ is the luminosity distance, $S_\gamma$ the $\gamma$--ray
fluence in the observed energy band, $k$ is the bolometric correction
factor needed to find the energy emitted in a fixed energy range
(here, 1--10000 keV) in the rest frame of the source.

We have used a different method, which enable us to weight the
multidimensional fit for the errors on the three independent variables
$E_{\rm \gamma, iso}$, $E^\prime_{\rm p}$ and $t^\prime_{\rm j}$. By
extending to the three dimensional space the procedure for the fit of
a straight line to data with errors on two coordinates (Press et
al. 1999) we use the $\chi^2$ statistics to find the best fit.

\begin{figure*}
\vskip -0.5 true cm
\centerline{\psfig{figure=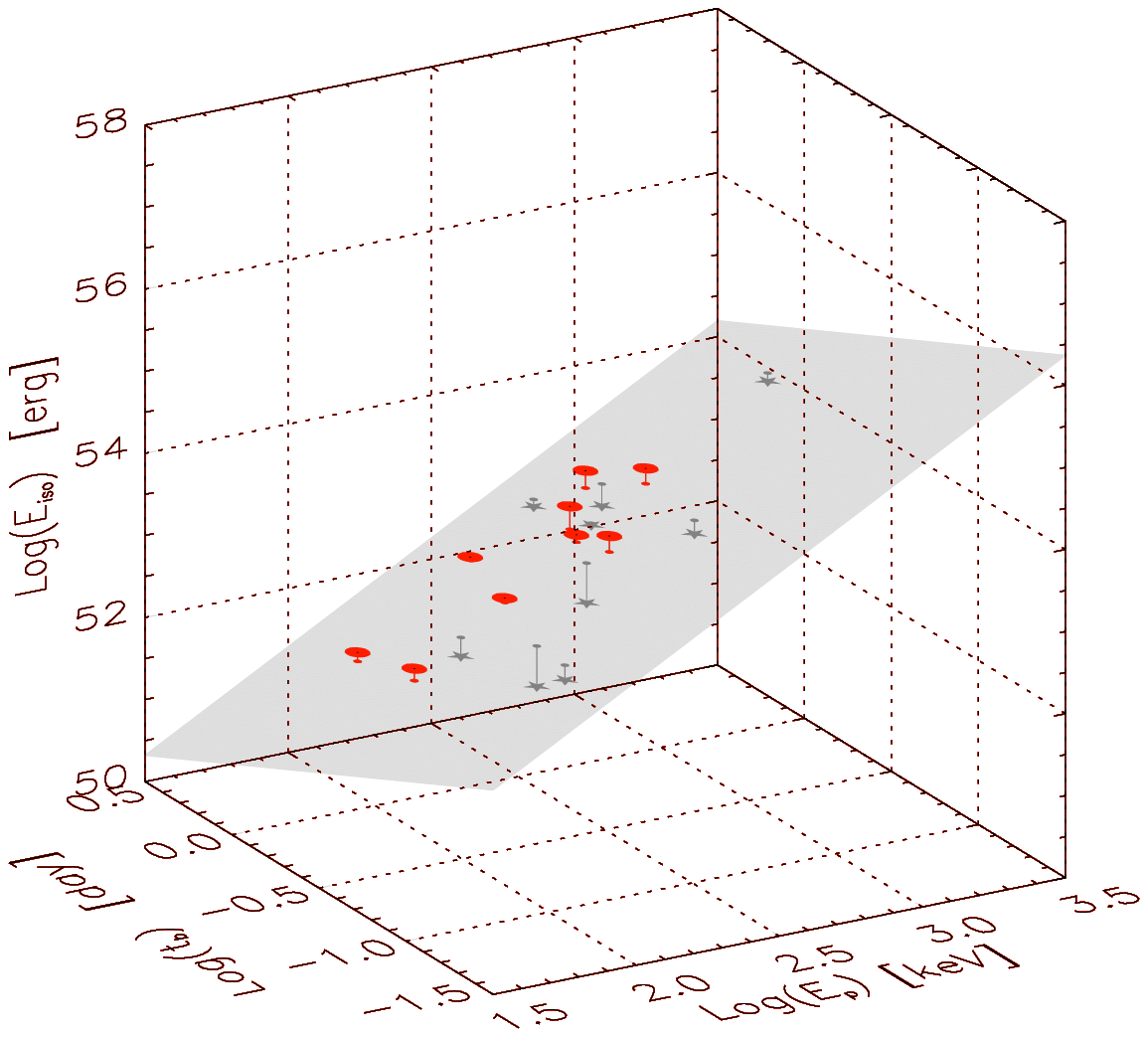,angle=0,width=16cm}}
\vskip -0.2 true cm
\caption{ Three dimensional representation of the $E_{\rm \gamma,
iso}(E^\prime_{\rm p}, t^\prime_{\rm j})$ correlation.  Data points
are from Tab.\ref{tabout}. The plane is the best fit to the data
points as represented by Eq. \ref{corr_fin} which has a reduced
$\chi^{2}_{\rm r}=1.49$.  The red points lies above the best fit plane
while the grey stars are below the plane. The height of each point
with respect to the best fit plane is also shown.}
\label{zhang}
\end{figure*}

\begin{figure}
\vskip -0.5 true cm
\centerline{\psfig{figure=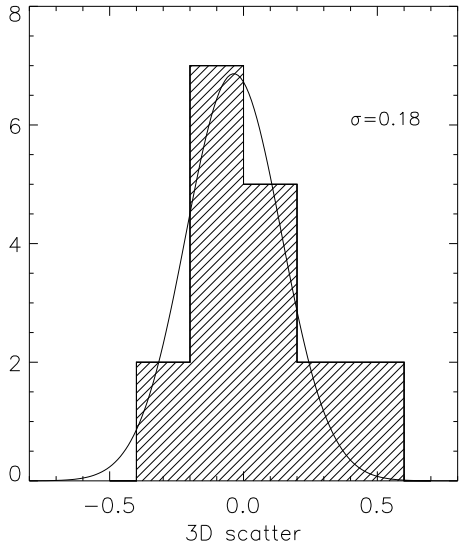,angle=0,width=8cm}}
\vskip -0.2 true cm
\caption{
Histogram of the scatter of the data points computed
perpendicular to the best fit plane in the 3D space of
Fig. \ref{zhang}. The solid line represents the gaussian fit which has
a $\sigma=0.18$.}
\label{scatt}
\end{figure}

Assuming a $\Omega_{\rm M}=0.3$ and $h=\Omega_\Lambda=0.7$ cosmology,
we find
\begin{equation}
E_{\rm \gamma,iso, 53} \, = (1.12\pm 0.11)\, 
\left(E^{\prime}_{\rm p} \over 295\, \rm keV \right)^{1.93\pm0.17} \,  
\left(t^{\prime}_{\rm j} \over 0.51\rm d\right)^{-1.08\pm0.17}
\label{corr_fin}
\end{equation}
with a reduced $\chi^{2}_{\rm r}=1.49$.
The 3D plot of the data points in the $E_{\rm \gamma, iso}$,
$E^\prime_{\rm p}$ and $t^\prime_{\rm j}$ space and the best fit plane
as defined by Eq. \ref{corr_fin} are represented in Fig. \ref{zhang}.
Similarly to what has been done in 2D (GGL04) we can define the
scatter of the data points around the best fit plane through their
distance computed perpendicular to this plane.  The histogram of the
scatter is reported in Fig.\ref{scatt} and, when fitted with a
gaussian it has a $\sigma=0.18$.


Comparing Eq. \ref{corr_fin} with the original result of LZ05 we
obtain a value of the $t^\prime_{\rm j}$ exponent closer to unity, but
still consistent with the value of LZ05 (which was also consistent with
unity, due to the relatively larger uncertainty).  As explained in the
next section, a value equal to unity is crucial to make the LZ05
correlation and the Ghirlanda correlation mutually consistent.

\begin{figure}
\vskip -0.5 true cm
\centerline{\psfig{figure=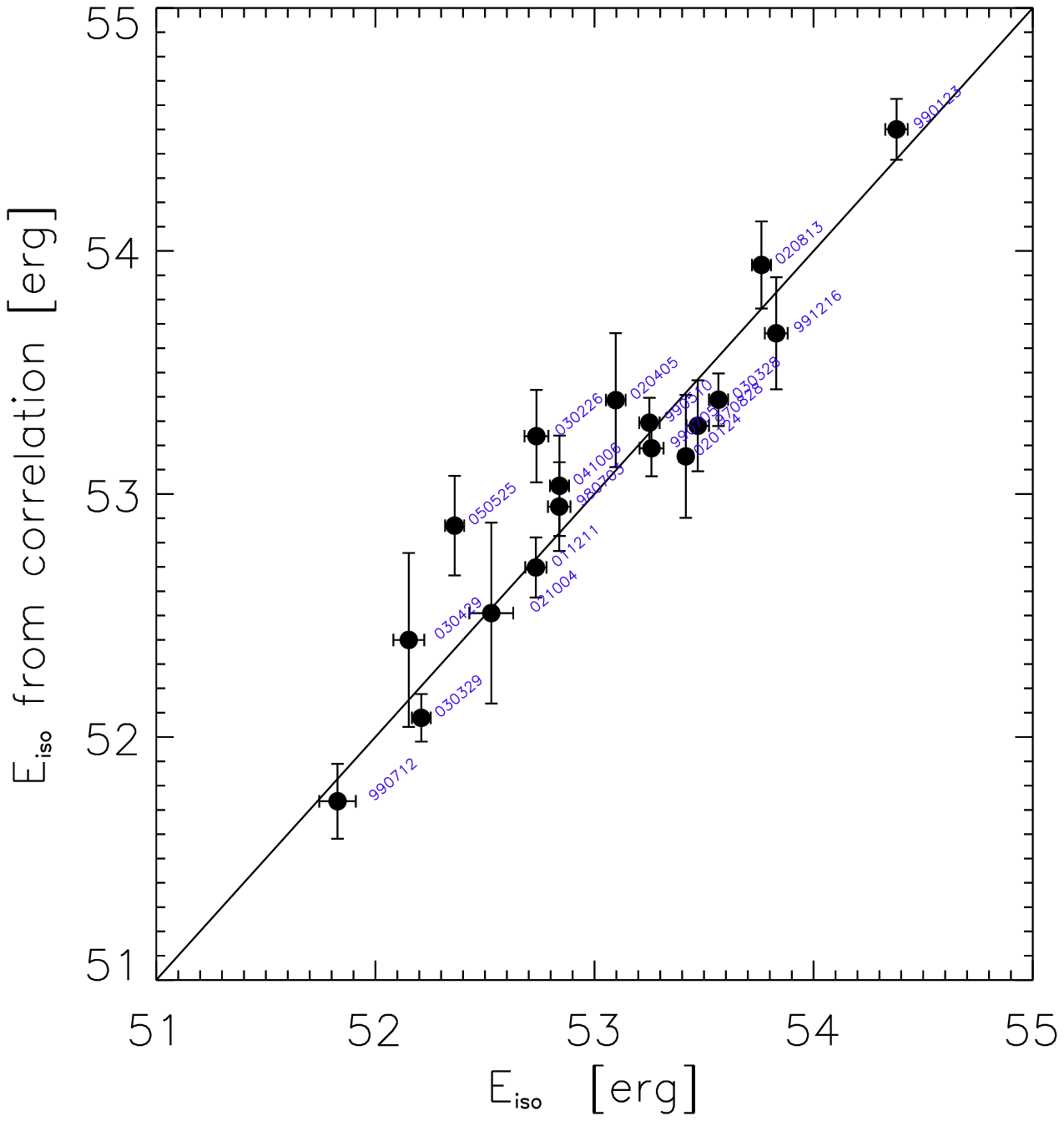,angle=0,width=8cm}}
\vskip -0.2 true cm
\caption{ 
$E_{\rm \gamma, iso}$ calculated through the best fit
multidimensional correlation $E_{\rm \gamma,iso}(E^\prime_{\rm p},
t^\prime_{\rm j})$ (Eq. \ref{corr_fin}) as a function of the same
quantity calculated through Eq. \ref{eiso}. }
\label{ee}
\end{figure}

We show in Fig. \ref{ee} the values of $E_{\rm \gamma,iso}$ calculated
through Eq. \ref{eiso} as a function of $E_{\rm \gamma,iso}$
calculated through the best fit of the correlation $E_{\rm\gamma,
iso}(E^\prime_{\rm p}, t^\prime_{\rm j})$ found using our data and our
method, and assuming a $\Omega_{\rm M}=0.3$ and $h=\Omega_\Lambda=0.7$
cosmology.  As can be seen, there is a very good agreement.

It is interesting, in view of the discussion of the following
sections, also to fit the $E_{\rm \gamma, iso}$, $E^\prime_{\rm p}$
and $t^\prime_{\rm j}$ correlation by forcing the slope of the
$t^\prime_{\rm j}$ to be --1.  Fixing this slope we find 
$E_{\rm \gamma, iso}\, t^\prime_{\rm j} \propto E_{\rm
p}^{1.91\pm0.1}$, with a reduced $\chi^2_r=1.51$.

\subsection{The updated Ghirlanda correlation}

\begin{figure*}
\vskip -0.5 true cm
\centerline{\psfig{figure=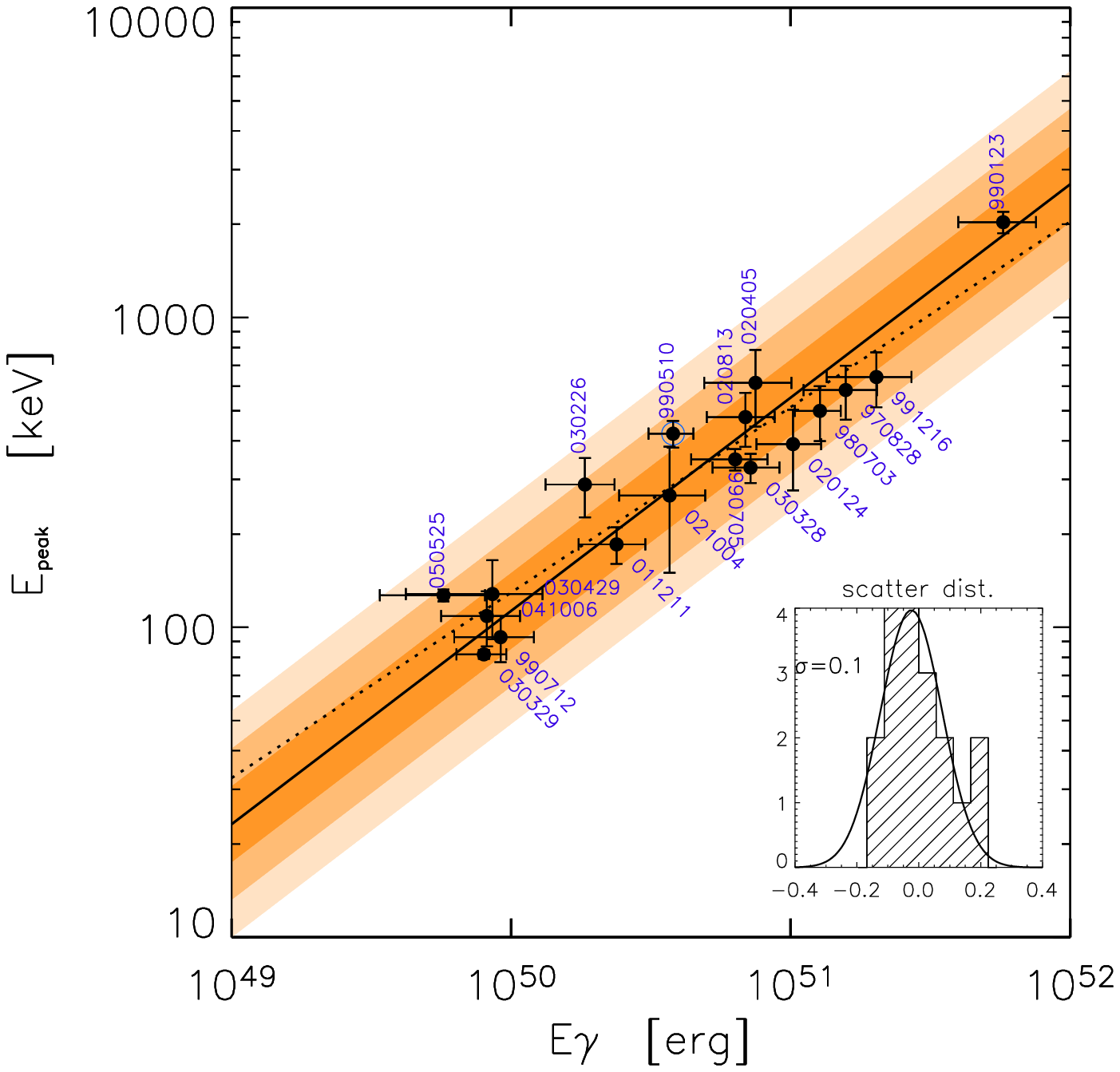,angle=0,width=16cm}}
\vskip -0.2 true cm
\caption{ The updated Ghirlanda correlation between the rest frame
spectral peak energy $E^\prime_{\rm p}$ and the collimation corrected
energy $E_{\gamma}$ as found with the 18 GRBs reported in
Tab. \ref{tabin}.  The solid line represents the best fit powerlaw
model obtained accounting for the errors on both coordinates
(Eq. \ref{ggl_new}) which has a reduced $\chi^{2}_{\rm r}=1.4$ (16
dof) and a slope of 0.69$\pm$0.04. We also show the fit obtained with
the simplest linear regression, i.e. without accounting for the errors
on the coordinates (dotted line, slope equal to 0.6).  The circled
point represents GRB~990510 which alone contributes to the 27\% of the
total $\chi^{2}$ of the fitted model.  The names of the 18 GRBs are
also reported.  The shaded areas represent the regions corresponding
to the 1, 2 and 3$\sigma$ scatter around the best fit correlation.
The {\it insert} reports the distribution (hatched histogram) of the
scatter of the data points computed perpendicularly to the best
correlation (solid line in main plot) and its gaussian fit (solid line
in the insert) which has a $\sigma=0.1$. }
\label{ghirla}
\end{figure*}

Using the same data listed in Tab. \ref{tabin} we calculate the
updated version of the Ghirlanda correlation. We find a Spearman
correlation coefficient $r_{s}=$0.93 with a chance probability
P=2.3$\times 10^{-8}$. We report in Fig. \ref{ghirla} the updated
correlation with the 18 GRBs reported in Tab. \ref{tabin}.

We fitted this correlation with a powerlaw model accounting for the
errors on both variables, i.e. $E^\prime_{\rm p}$ and $E_{\gamma}$
(using the routine {\it fitexy} of Press et al. 1999).  For
$\Omega_{\rm M}=0.3$, $\Omega_\Lambda=h=0.7$ we find
\begin{equation}
\left({E^\prime_{\rm p} \over 100\, {\rm keV}}\right) \, =\, 
(2.79\pm0.15)\, 
\left({E_\gamma\over 2.72 \times 10^{50}\, {\rm erg}}\right)^{0.69\pm 0.04}
\label{ggl_new}
\end{equation}
with a reduced $\chi^{2}_{\rm r}=1.4$ for 16 degrees of freedom.  
The errors on its slope and normalization are calculated in the
``barycenter" of $E^\prime_{\rm p}$ and $E_{\gamma}$, where the slope
and normalization errors are uncorrelated (Press et al. 1999).  
We also note that the simplest linear regression fit (i.e. without
accounting for errors on the variables) gives a slope of 0.6 (dotted
line in Fig. \ref{ghirla}).

Fig. \ref{ghirla} shows the correlation for the 18 GRBs and its best
fit represented by Eq. \ref{ggl_new} (solid line).  This updated
correlation has a slope which is consistent with the original value
found in GGL04.  We also computed the scatter of the data points
around this correlation (insert of Fig.\ref{ghirla}).  This scatter is
defined as the distance in the $\log E^\prime_{\rm p} - \log
E_{\gamma}$ plane of each data point from the best fit correlation.
We find that if fitted with a gaussian its standard deviation is
$\sigma=0.1$, i.e. lower than the value originally found by GGL04.

For completeness, we also computed the Amati correlation with the 18
GRBs reported in Tab. \ref{tabin}.  By weighting for the errors on
$E_{\rm p}$ and $E_{\gamma, \rm iso}$, we find a relatively poor fit
with a reduced $\chi^{2}_{\rm r}=5.22$ and a best fit correlation
$E_{\gamma,\rm iso} \propto E^{0.57\pm 0.02}_{\rm p}$.

\subsection{The Ghirlanda correlation in the case of a wind density profile}

\begin{figure*}
\vskip -0.5 true cm
\centerline{\psfig{figure=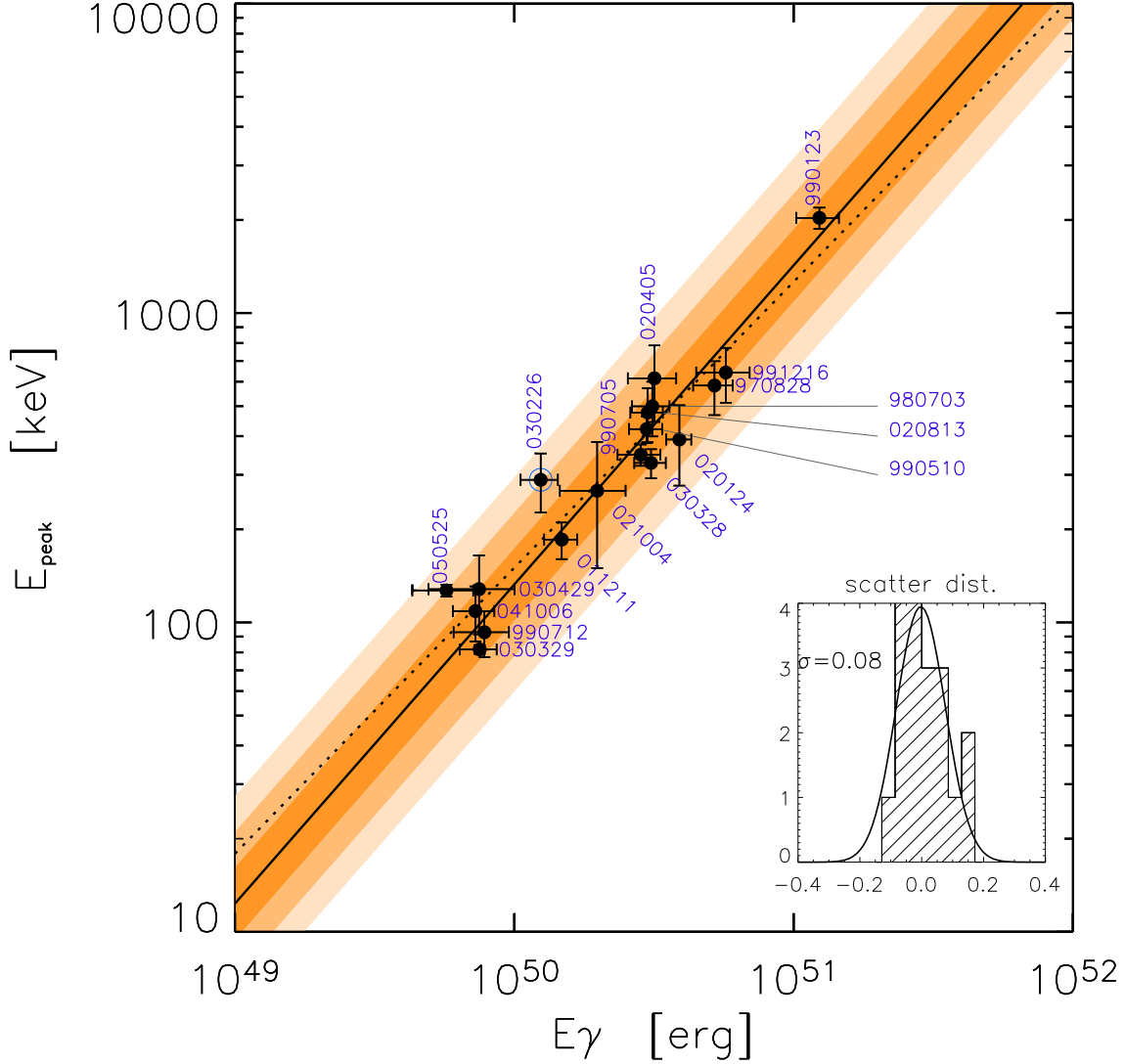,angle=0,width=16cm}}
\vskip -0.2 true cm
\caption{ The Ghirlanda correlation in the case of a wind profile of
the external medium density as found with the 18 GRBs reported in
Tab. \ref{tabin} (col. 5).  The values of $E_{\gamma}$ are reported in
Tab. \ref{tabout}.  The solid line represents the best fit powerlaw
model obtained accounting for the errors on both coordinates
(Eq. \ref{gglw}) which has a reduced $\chi^{2}_{\rm r}=1.125$ (16 dof)
and a slope of 1.09$\pm$0.06.  We also show the fit obtained with the
simplest linear regression, i.e. without accounting for the errors on
the coordinates (dotted line), which has a slope of 0.92.  The circled
point represents GRB 030326 which is giving the largest contribution
(23\%) to the best fit reduced $\chi^2$.  The shaded regions represent
the 1, 2 and 3$\sigma$ scatter around the best fit correlation.  The
names of the 18 GRBs are indicated.  The {\it insert} reports the
distribution (hatched histogram) of the scatter of the data points
computed perpendicularly to the best correlation (solid line in the
main plot) and its gaussian fit (solid line in the insert) which has a
$\sigma=0.08$.  }
\label{ghirla_wind}
\end{figure*}

If the external medium is distributed with an $r^{-2}$ density profile
the semiaperture angle of the jet is related to the achromatic jet
break through (Chevalier \& Li 2000):
\begin{equation}
\theta_{\rm j, w}\, = \, 0.2016 \, 
\left( {t_{\rm j,d} \over 1+z}\right)^{1/4} 
\left( { \eta_\gamma\ A_* \over E_{\rm \gamma,iso,52}}\right)^{1/4}
\label{thetaw} 
\end{equation}
where we assume $n(r)=Ar^{-2}$ and $A_*$ is the value of $A$ ($A=\dot
M_{\rm w} /(4\pi v_{\rm w})=5\times 10^{11}A_*$ g cm$^{-1}$ ) when
setting the mass loss rate due to the wind $\dot M_{\rm w} =10^{-5}
M_\odot$ yr$^{-1}$ and the wind velocity $v_{\rm w}=10^3$ km s$^{-1}$,
according to the Wolf--Rayet wind physical conditions.

In the wind case we use Eq. \ref{thetaw} to correct the isotropic
energies $E_{\rm \gamma, iso}$ by the factor $(1-\cos\theta_{\rm
j,w})$.  Given the few still uncertain estimates of the $A_*$
parameter, we assume the typical value (i.e. $A_*=1$) for all bursts
neglecting for the moment the possible uncertainty on this parameter.

In the wind case we find a correlation
between $E^\prime_{\rm p}$ and $E_\gamma$ with a Spearman rank 
correlation coefficient $r_{\rm s}=0.92$ ($P=6.9\times 10^{-8}$). 
The fit with a powerlaw model gives
\begin{equation}
{E^\prime_{\rm p} \over 100\, {\rm keV}}\, =\, 
(3.0\pm0.16)\, \left( {E_\gamma \over 2.2\times 10^{50}\, 
{\rm erg}}\right)^{1.03\pm 0.06}
\label{gglw}
\end{equation}
with a reduced $\chi^{2}_{\rm r}=1.13$ for 16 dof (see
Fig. \ref{ghirla_wind}).  Note that the exponent of this new relation
is entirely consistent with unity.  The scatter of the points (insert
of Fig. \ref{ghirla_wind}) around the best fit correlation is fitted
by a gaussian with $\sigma=0.08$.

Since we have no knowledge of the uncertainty associated with the
$\eta A_*$ parameter entering in Eq. \ref{thetaw}, we estimate that,
for the assumed typical value $\eta A_*=0.2$, an error $\sigma_{\eta
A_*}\le 20$\% does not dominate the fit of the correlation (i.e. the
reduced $\chi^{2}_{\rm r}$ is not much smaller than 1).

\begin{figure}
\vskip -0.5 true cm
\centerline{\psfig{figure=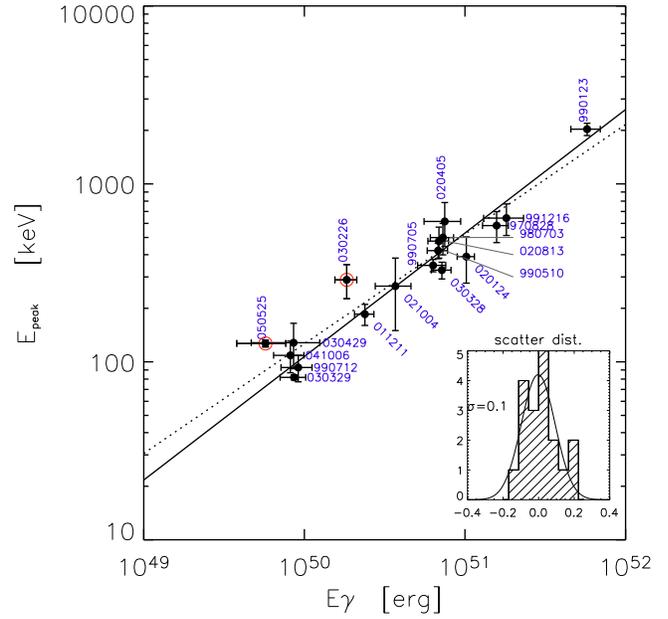,angle=0,width=9cm}}
\vskip -0.2 true cm
\caption{
The Ghirlanda correlation in the case of a homogeneous
external medium assuming a universal value for the density, i.e n=3.0
without uncertainty, for comparison with the wind case. The solid line
represent the best fit powerlaw model obtained accounting for the
errors on both coordinates which has a reduced $\chi^{2}_{\rm r}=1.4$
(16 dof) and a slope of 0.69$\pm 0.04$.}
\label{ghirla_ncost}
\end{figure}

Since we have assumed that all the $A_*$ values are equal and have no
errors, the resulting $\chi_r^2$ of the wind case should be compared
with the the case of homogeneous density assuming all the $n$ values
equal (we set $n=3$ cm$^{-3}$) with no error (see Tab. \ref{tabout}).
This case is shown in Fig. \ref{ghirla_ncost}.  In this case we obtain
$\chi_r^2=1.4$ to be compared with the $\chi_r^2=1.125$ of the wind
case.  We conclude that the wind case gives a somewhat better
$\chi_r^2$ and a somewhat tighter (smaller scatter) and a
steeper correlation than the correlation found in the case of an
homogeneous medium. Therefore, even if the wind density profile
is not favored by the afterglow model we cannot discard it on the basis
of the Ghirlanda relation.

\begin{figure}
\vskip -0.5 true cm
\centerline{\psfig{figure=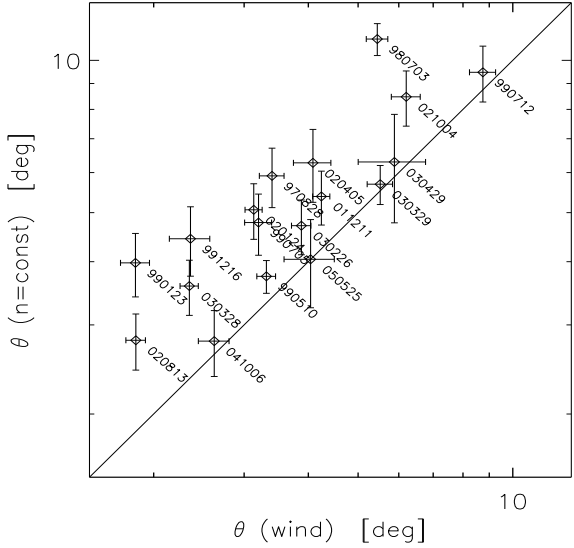,angle=0,width=9cm,height=8cm}}
\vskip -0.2 true cm
\caption{Jet opening angles calculated for a homogeneous density
profile ($\theta_{j}$) and for a wind density profile ($\theta_{j,w}$)
for the 18 GRBs reported in Tab. 2 (col. 4 and 7 respectively). 
}
\label{ang}
\end{figure}

The jet opening angles calculated in the case of a homogeneous medium
or in the case of a wind density profile, for the sample of 18 GRBs,
is reported in Fig. \ref{ang}. The angle calculated in the wind case
(Eq. \ref{thetaw}) is sistematically smaller than in the homogeneous
medium case (Eq. \ref{theta}).

\section{Consistency of the empirical correlation with the model 
dependent correlations}

In this section we will demonstrate that the Ghirlanda correlation
either assuming an homogeneous circumburst density or a wind density
profile and the LZ05 correlation are mutually consistent.  This
allows us to make some interesting considerations on the scatter of
the Ghirlanda correlations. The fact that the LZ05 and the Ghirlanda
correlation are mutually consistent have been already pointed out by
LZ05 (see also Xu 2005) in the case of an homogeneous medium.

For the simple analytical demonstration, we will consider a generic
form of the LZ05 correlation, namely:
\begin{equation}
E_{\rm \gamma, iso}\, \propto E_{\rm p}^{\prime A} t^{\prime B}_{\rm j}
\label{ab}
\end{equation}

\subsection{Homogeneous density}

Adopting the standard fireball scenario, assuming a uniform jet and an
homogeneous circumburst density distribution, the relation between
$t^\prime_{\rm j}$ and $\theta_{\rm j}$ is given by Eq. \ref{theta}.
Inserting it into Eq. \ref{ab} one obtains
\begin{equation}
E_{\rm \gamma,iso}\theta_{\rm j}^2 \sim E_\gamma \propto
E^{\prime 3A/(3-B)}_{\rm p}\ 
\theta_{\rm j}^{6(B+1)/(3-B)} (n\eta_\gamma)^{-B/(3-B)}
\label{5d}
\end{equation}
where we have used the small angle approximation $(1-\cos\theta_{\rm
j}) \propto \theta_{\rm j}^2$.  Eq. \ref{5d} relates five variables.

We have shown that the Ghirlanda correlation is characterized by
a scatter $\sigma\sim 0.1$ (Sec.~3.2 and Fig.~\ref{ghirla}). If this
scatter is entirely due to a dispersion of the $n\eta_{\gamma}$ values
(and very likely also to the errors of measurements on the
observables) and not to   $\theta_{\rm j}$, we derive the condition
$B=-1$ and Eq.~\ref{5d} reduces to a relation between $E_\gamma$ and
$E^\prime_{\rm p}$. This value of $B$ is consistent with the value
found from the fit of the LZ05 correlation (i.e. $B=-1.08\pm0.17$)
with the 18 GRBs of our sample. If the exponent $B=-1$, then the slope
$g$ of the Ghirlanda correlation (i.e. $E^\prime_{\rm p} \propto
E_\gamma^g$) is related to the exponent $A$ of the LZ05 correlation
through $g= 4/(3A)$.

\begin{figure}
\vskip -0.5 true cm
\centerline{\psfig{figure=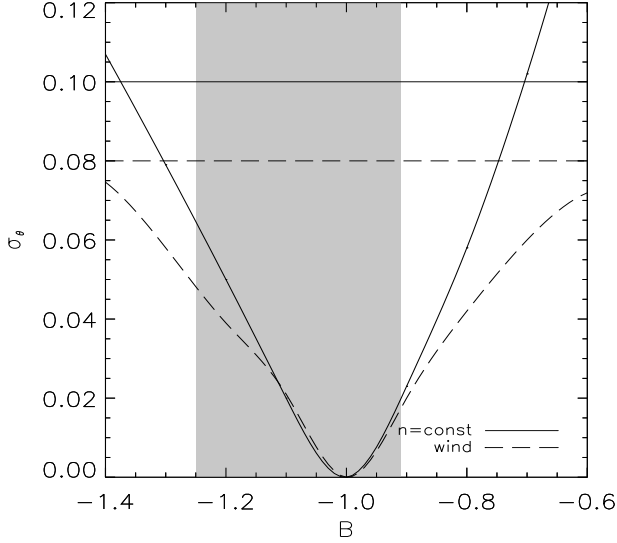,angle=0,width=9cm,height=8cm}}
\vskip -0.2 true cm
\caption{ Standard deviation of the distribution of $\theta^h(B)$
where $h(B)$ is given by Eq. \ref{5d} and Eq. \ref{5dw} for the
homogeneous (solid line and filled points) and wind density profile
(long dashed line and open circles) as a function of the parameter
$B$. The horizontal lines represent the scatter of the Ghirlanda
correlation in the two scenarios (solid and long--dashed line
respectively). The shaded region represents the ($1\sigma$)
uncertainty on the parameter $B$ (=$-1.08\pm$0.17) found through the
fit of the LZ05 correlation with the 18 GRBs.}
\label{sigma_b}
\end{figure}

On the other extreme, if the scatter of the Ghirlanda correlation
is completely due to $\theta_{\rm j}$ (and to the errors of
measurements on the observables) we may still derive a range of
allowance for the parameter $B$. We computed the standard deviation
$\sigma$ of the distribution of $\theta_{\rm j}^{6(B+1)/(3-B)}$ as a
function of $B$. This is represented in Fig.~\ref{sigma_b} by the
solid line. If we compare $\sigma$ with the scatter of the Ghirlanda
correlation (in the homogeneous density case - solid horizontal line
in Fig.~\ref{sigma_b}) we can find quite shallow constraints on
$B\in(-1.3,-0.7)$.

Clearly, the intermediate case corresponds to both $n\eta_{\gamma}$
and $\theta_{\rm j}$ contributing to the scatter. If we consider the
range of $B$ as found by fitting the LZ05 correlation
(Eq.~\ref{corr_fin}), i.e. $B=-1.08\pm0.17$ (shaded region in
Fig.~\ref{sigma_b}) then the term $\theta_{\rm j}^{6(B+1)/(3-B)}$ can
contribute at most for the $\sim 70$\% of the total scatter of the
Ghirlanda correlation.

\subsection{Wind density profiles}

If $t_{\rm j}$ is related to the semiaperture angle of the jet
according to Eq. \ref{thetaw}, the LZ05 correlation implies:
\begin{equation}
E_{\rm \gamma,iso}\theta_{\rm j,w}^2 \sim E_\gamma \propto
E^{\prime A/(1-B)}_{\rm p}\ 
\theta_{\rm j}^{2(B+1)/(1-B)} (n\eta_\gamma)^{-B/(1-B)}
\label{5dw}
\end{equation}
For the same considerations reported in the previous section, we
derive $B=-1$ if only the term $n\eta_\gamma$ contributes to the
scatter of the Ghirlanda correlation in the wind density case
(i.e. $\sigma\sim0.08$, Sec.~3.3 and Fig.\ref{ghirla_wind}). In this
case the relation between the Ghirlanda correlation and the LZ05
correlation implies that $E^\prime_{\rm p}\propto E_\gamma^{2/A}$.

If we consider that the scatter of the Ghirlanda correlation
(long dashed horizontal line in Fig.~\ref{sigma_b}) is entirely due to
the term $\theta_{\rm j}^{2(B+1)/(1-B)}$ we can derive even shallower
(with respect to the homogeneous case) constraints on the parameter
$B\in(-1.45,-0.65)$ (Fig.~\ref{sigma_b} - long dashed curve). Again in
the intermediate case, i.e. both $n\eta_\gamma$ and $\theta_{\rm j}$
contributing to the scatter of the Ghirlanda correlation, if we assume
the range of possible values of $B$ resulting from the fit of the LZ05
correlation than the term $\theta_{\rm j}^{2(B+1)/(1-B)}$ contributes
at most for the $\sim$60\% to the scatter observed in the Ghirlanda
correlation.

\section{The Ghirlanda correlation in the comoving frame}

The values of $E^\prime_{\rm p}$ and $E_\gamma$ we derive are the ones
seen in the GRB rest frame but not in the fireball comoving frame.

In the simplest and standard scenario, which assumes that the
observer's line of sight is within the jet opening angle and that the
jet is homogeneous, $E_{\rm p}$ and $E_{\gamma}$ are both boosted by a
factor $\sim 2\Gamma$, i.e.  the blueshift of the photons.
 
Then the comoving $E^{\rm com}_\gamma = E_\gamma/(2\Gamma)$ and
$E^{\rm com}_{\rm p}=E^\prime_{\rm p}/(2\Gamma)$.  Here $\Gamma$ is
the bulk Lorentz factor of the fireball emitting the prompt radiation.

We then exploit the fact that we see a very tight correlation (in the
rest, but not comoving, frame) to pose some limits on the physics
(i.e. the dynamics and the radiation process) of bursts.

Let us assume that the most general Ghirlanda correlation
described by a generic power law, i.e.  $E^\prime_{\rm p} \propto
E_\gamma^g$, which can then represents the correlation for both
the homogeneous and wind density profiles.  In the comoving frame,
each point must be corrected by the bulk Lorentz factor of that burst.
If these bulk Lorentz factors are uncorrelated with $E_\gamma$ or
$E^\prime_{\rm p}$, then in the comoving frame the correlation is
destroyed.  It would then seem very strange that two quantities that
are not correlated in the comoving frame appear to be correlated in
the rest frame.  We then are obliged to assume that $\Gamma$ is a
function of $E_\gamma$ (or, equivalently, of $E^{\rm com}_\gamma$).
To this aim, let assume a simple power law relation:
\begin{equation}
\
\Gamma \,\propto \, \left( E^{\rm com}_\gamma \right)^x; \qquad
\Gamma \, \propto\,  E_\gamma^{x/(1+x)}; \quad  
\end{equation}
The two relations above are equivalent.
In the comoving frame we have:
\begin{equation}
E^{\rm com}_{\rm p} \, \propto 
\left( E^{\rm com}_\gamma \right)^{xg-x+g}
\label{corr_com}
\end{equation}
Note that:
\begin{itemize}
\item
If $x=0$ (i.e. all bursts have the same Lorentz factor) then the slope
in the comoving and in the rest frame is the same;
\item
If $g=1$ (i.e. wind case), the correlation is linear also in 
the comoving frame;
\item
The exponent $x$ can be thought as determined by a particular
dynamical model of the fireball.  Once $x$ is fixed, then the exponent
$(xg-x+g)$ appearing in Eq. \ref{corr_com} should be explained by the
radiative process.  Viceversa, if we have reason to fix, through a
specific radiation model, the exponent of Eq. \ref{corr_com}, then we
have information on the dynamics of the fireball.
\end{itemize}

We stress the fact that for the wind case (i.e. $g=1$) we obtain {\it
a linear relation, whose slope is therefore ``Lorentz invariant"}.
This of course would greatly help to explain the existence of the
Ghirlanda correlation, since one of the main parameters, the bulk
Lorentz factors, does not enter to determine it.  In other words, in
the case of the wind, the explanation of the correlation should be
found in the radiation process only, independently of the dynamics.

\section{Discussion}

We have shown that the model--independent correlation recently found
by LZ05 between the isotropic emitted energy $E_{\rm \gamma, iso}$,
the rest frame peak energy $ E^\prime_{\rm p}$ and the jet break time
$t^\prime_{\rm j}$, calculated in the rest frame, is equivalent to the
Ghirlanda correlation.

We have also shown that an even tighter correlation is found assuming
that the circumburst density is distributed with an $r^{-2}$ wind
profile.  Remarkably enough, this Ghirlanda--wind correlation is
linear, and this slope is independent of the Lorentz correction needed
to find out the correlation in the comoving frame.

We are aware that the wind--like distribution of the circumburst
medium is not favored by the existing afterglow modeling, but the
advantage of having a linear Ghirlanda correlation is so great to
justify a deeper analysis, to see if there can be some neglected
effects which might be able to hide the presence of the wind.  This is
however out of the scope of the present paper, and we defer this issue
to future studies.  Here we only mention that one of the main
assumption of the afterglow modeling might be particularly suspect,
namely the hypothesis that the equipartition parameters $\epsilon_B$
and $\epsilon_e$ (i.e. the fraction of the dissipated energy converted
in the magnetic field and in the electron energy, respectively) are
kept fixed during the entire afterglow phase (while they have very
different values from burst to burst).  For instance, in the case of
$\epsilon_B \propto \Gamma^{-\lambda}$, the synchrotron cooling is
enhanced at later times with respect to the case of a constant
$\epsilon_B$, and this makes the light curve of a fireball moving in a
wind circumburst environment to mimic the evolution of a fireball
expanding in a uniform medium with a constant $\epsilon_B$ (the two
cases becomes almost indistinguishable for $\lambda=2$).  Consider
also that all the pieces of evidence we have up to now point towards a
massive stellar progenitor of GRBs, and it is difficult to understand
why there is no sign of winds around such massive stars.  The fact
that $\epsilon_B$ changes (as long as its evolution law does not
change), does produces a different (with respect to a not--evolving
$\epsilon_B$) decay law of the afterglow flux, but with no breaks.
Therefore in this case $\epsilon_B$ does not enter in the estimate of
the jet opening angle (as long as we are in the adiabatic regime).

One interesting developement (Ghirlanda et al. 2005a) of having
found a somewhat tighter and steeper Ghirlanda correlation in the wind
density case (with respect to the homogenous case) is to use it to
constrain the cosmological prameters similarly to what already done
through the Ghirlanda correlation in the homogeneous case (Ghirlanda
et al. 2004a; Firmani et al. 2005).

As pointed out by Ghirlanda, Ghisellini \& Firmani (2005), the
existence of the Ghirlanda correlation explains the Amati correlation
between $E_{\gamma \rm iso}$ and $E^\prime_{\rm p}$ (Amati et
al. 2002, Ghirlanda, Ghisellini \& Lazzati 2004), and in particular
explains why the Amati correlation has a much larger scatter than the
Ghirlanda correlation.  In fact if GRBs are characterized by a
distribution of semiaperture angles for each value of $E^\prime_{\rm
p}$, then one sees a variety of $E_{\rm \gamma,iso}$--values for each
value of $E^\prime_{\rm p}$.  If the (observed) distribution of
aperture angles turns out to have a preferred value, where it peaks,
then this naturally produces a correlation in the $E^\prime_{\rm
p}$--$E_{\rm \gamma,iso}$ plane.  Furthermore, it is conceivable that
the bursts with spectroscopically measured redshifts are the
brightest, hence with the smallest aperture angles (for a given
redshift), hence lying at the large $E_{\rm \gamma, iso}$ end of the
real distribution.  We have evidence that this is just what is
happening (Ghirlanda, Ghisellini \& Firmani 2005).

Therefore the Amati correlation can be easily explained by assuming
the existence of i) the Ghirlanda correlation and ii) the existence of
a (peaked) distribution of jet aperture angles.  Thus what remains to
be explained is the Ghirlanda correlation itself which is relating the
intrinsic collimation corrected quantities.  We have pointed out that
there might be a difference between the apparent and the comoving
Ghirlanda correlation, and from the theoretical point of view it is
the comoving one that should be explained.  As Rees \& Meszaros (2005)
pointed out, a tight relation between the peak energy of the spectrum
and the total emitted energy reminds of a thermal process, where the
peak energy is a measure of a temperature.  It is in this direction
that we plan to investigate in the future.

Finally, let us comment, regarding the linear, wind--like, Ghirlanda
correlation, which implies that the number of ``relevant" photons (the
ones with energies close to $E^\prime_{\rm p}$) is constant in all
bursts, and approximately equal to $N_\gamma=10^{57}$, a number
(coincidentally?) close to the number of baryons in one solar mass.
In the ``standard" scenario, in which the primary energy is injected
close to the putative newly born black hole in a high entropy form,
the requirement of a fixed number of photons $N_\gamma$ translates in
the requirement that the product of the injected energy $E$ and the
typical size of the injection region $R$ is constant.  This can be
seen in a simple way by noting that $a T^4 R^3/(kT) \sim N_\gamma$,
where $T$ is the temperature of the initial blackbody.  Since $E\sim
kT N_\gamma$, one arrives to $ER\propto N_\gamma^{4/3}=$const.  In the
framework of the standard scenario, in the absence of additional
injection of energy, the number of photons is conserved during the
acceleration and the coasting phase.  Then, when the fireball becomes
transparent, this blackbody--like component has a fixed number of
photons.  But bursts with different baryon loading would become
transparent at different times, meaning that different fractions of
the energy initially contained in the radiation field have survived to
the conversion to the bulk kinetic energy.  This means different
efficiency factors $\eta_\gamma$.  And this is contrary to one of the
main assumptions leading to the construction of the Ghirlanda
correlation itself, which has been derived assuming the same
$\eta_\gamma$ for all bursts.

If some extra dissipation of kinetic energy occurs after the acceleration 
phase, before the transparency radius, and the main radiation process is 
Comptonization, then we can reconvert part of the kinetic energy into 
radiation energy leaving the number of photons unaltered.
The available time to do it will be proportional to the transparency
radius, so we have the positive feedback that sources which becomes
transparent later (i.e. more baryon loaded, and hence with a 
radiation content which will be less energetic), have more possibilities to 
be re--energized by this putative dissipation (Rees \& Meszaros 2005).
This positive feedback can narrow the range of radiative efficiencies.
In any case, we face the problem to explain why bursts with the
same number of photons have different total energies.

Apart from these theoretical considerations, there might be an
``observational" way to see if the real $E_\gamma$--$E^\prime_{\rm p}$
is wind--like or constant density--like.  Assume in fact that there is
indeed a correlation between $E_{\rm p}$ and $E_{\gamma}$ and suppose
that it will be possible to estimate the bulk Lorentz factor of the
fireball during the emission of the prompt (that should be roughly
equal to the bulk Lorentz factor of the very early afterglow, and
controls the time at which there is the peak flux of the afterglow
itself).  Preliminary attempts to estimate it have already been done
by e.g.  Soderberg \& Ramirez--Ruiz (2002) for GRB 990123.  Then, if
by ``de--beaming" the $E_\gamma$ and $E^\prime_{\rm p}$ values
(i.e. computing them in the comoving frame) in the case of an
homogeneous density one finds an equally tight correlation, this would
strongly point in favor of the validity of the homogeneous density
hypothesis.  Perhaps more importantly, one will also find the relation
between $\Gamma$ and $E^{\rm com}_\gamma$.  On the contrary, if the
de--beamed quantities (calculated in the case of an homogeneous
density) do not correlate, this will strongly argue in favor of the
wind hypothesis (in such a case the de--beamed quantities still
correlate).

\begin{acknowledgements}
We thank the referee Amir Levinson for his valuable comments. We thank
Annalisa Celotti and Davide Lazzati for discussions.  The Italian MIUR
and INAF are thanked for founding (Cofin grant 2003020775\_002).

\end{acknowledgements}

\end{document}